\documentclass[twocolumn,preprintnumbers,amsmath,amssymb,superscriptaddress]{revtex4}

\usepackage{graphicx}
\begin{document}
\bibliographystyle{prsty}
\title{Emergence of quantum critical behavior in metallic quantum-well states of strongly correlated oxides
}
%\pdfoutput=1

\author{Masaki Kobayashi}
\altaffiliation{Author to whom correspondence should be addressed; masakik@post.kek.jp}
\affiliation{Photon Factory, Institute of Materials Structure Science, 
High Energy Accelerator Research Organization (KEK), 
1-1 Oho, Tsukuba 305-0801, Japan}
\author{Kohei Yoshimatsu}
\altaffiliation{Present address: Department of Applied Chemistry, Tokyo Institute of Technology, 
Tokyo 152-8522, Japan}
\affiliation{Photon Factory, Institute of Materials Structure Science, 
High Energy Accelerator Research Organization (KEK), 
1-1 Oho, Tsukuba 305-0801, Japan}
\affiliation{Department of Physics, University of Tokyo, 
7-3-1 Hongo, Bunkyo-ku, Tokyo 113-0033, Japan}
\author{Miho Kitamura}
\affiliation{Photon Factory, Institute of Materials Structure Science, 
High Energy Accelerator Research Organization (KEK), 
1-1 Oho, Tsukuba 305-0801, Japan}
\author{Enju Sakai}
\affiliation{Photon Factory, Institute of Materials Structure Science, 
High Energy Accelerator Research Organization (KEK), 
1-1 Oho, Tsukuba 305-0801, Japan}
\author{Ryu Yukawa}
\affiliation{Photon Factory, Institute of Materials Structure Science, 
High Energy Accelerator Research Organization (KEK), 
1-1 Oho, Tsukuba 305-0801, Japan}
\author{Makoto Minohara}
\affiliation{Photon Factory, Institute of Materials Structure Science, 
High Energy Accelerator Research Organization (KEK), 
1-1 Oho, Tsukuba 305-0801, Japan}
\author{Atsushi Fujimori}
\affiliation{Department of Physics, University of Tokyo, 
7-3-1 Hongo, Bunkyo-ku, Tokyo 113-0033, Japan}
\author{Koji Horiba}
\affiliation{Photon Factory, Institute of Materials Structure Science, 
High Energy Accelerator Research Organization (KEK), 
1-1 Oho, Tsukuba 305-0801, Japan}
\author{Hiroshi Kumigashira}
\altaffiliation{Author to whom correspondence should be addressed; hkumi@post.kek.jp}
\affiliation{Photon Factory, Institute of Materials Structure Science, 
High Energy Accelerator Research Organization (KEK), 
1-1 Oho, Tsukuba 305-0801, Japan}

\date{\today}

\begin{abstract}
Controlling quantum critical phenomena in strongly correlated electron systems, which emerge in the neighborhood of a quantum phase transition, is a major challenge in modern condensed matter physics.  
Quantum critical phenomena are generated from the delicate balance between long-range order and its quantum fluctuation.  
So far, the nature of quantum phase transitions has been investigated by changing a limited number of external parameters such as pressure and magnetic field.  
We propose a new approach for investigating quantum criticality by changing the strength of quantum fluctuation that is controlled by the dimensional crossover in metallic quantum well (QW) structures of strongly correlated oxides.  
With reducing layer thickness to the critical thickness of metal-insulator transition, crossover from a Fermi liquid to a non-Fermi liquid has clearly been observed in the metallic QW of SrVO$_3$ by \textit{in situ} angle-resolved photoemission spectroscopy.  
Non-Fermi liquid behavior with the critical exponent ${\alpha} = 1$ is found to emerge in the two-dimensional limit of the metallic QW states, indicating that a quantum critical point exists in the neighborhood of the thickness-dependent Mott transition.  
These results suggest that artificial QW structures provide a unique platform for investigating novel quantum phenomena in strongly correlated oxides in a controllable fashion.
\end{abstract}

%\pacs{73.21.Fg, 71.27.+a, 79.60.-i}

\maketitle

%\section{Introduction}
Dimensionality is one of the key parameters used for controlling the extraordinary physical properties of strongly correlated electron systems \cite{RMP_98_Imada, Nature_02_Valla, PRB_91_Nozaki, PRL_08_Moon, PRB_11_Malvestuto, Science_10_Shishido, Science_11_Boris, NatNanotech_14_King, SciAdv_15_Mikheev, NatMater_12_Hwang, Science_00_Sachdev}.  
The lowering of the dimensionality induces a variety of essential changes in the electronic properties and changes the complex interactions among the spin, charge, and orbital degrees of freedom of correlated electrons \cite{RMP_98_Imada, Nature_02_Valla, PRB_91_Nozaki, PRL_08_Moon, PRB_11_Malvestuto, Science_10_Shishido, Science_11_Boris, NatNanotech_14_King, SciAdv_15_Mikheev, NatMater_12_Hwang}.  
The dimensional crossover from three dimensions (3D) to two dimensions (2D) offers a privileged position for studying quantum critical phenomena \cite{Science_00_Sachdev}.  
In 3D systems, low-energy electronic states behave as quasiparticles (QPs) owing to the weakness of the quantum fluctuation of the order parameter.  Consequently, the physical properties are well described in the framework of Fermi liquid (FL) theory or by long-range ordered states with order parameters.  
In contrast, the quantum fluctuation is so strong in one-dimensional systems that even infinitely weak interactions break the QPs into collective excitations, and long-range order is prevented.  
In 2D system, the delicate balance between long-range order and quantum fluctuation hosts interesting quantum critical phenomena \cite{Science_00_Sachdev, RPP_03_Moriya, RMP_07_Lohneysen}.  
Therefore, the dimensional crossover from 3D to 2D is an ideal platform for systematically studying the quantum critical phenomena that emerge in the neighborhood of the quantum phase transition by utilizing the enhancement of the quantum fluctuation driven by dimensional crossover.

The likeliest system in bulk materials for 3D-to-2D control is layered complex oxides such as the Ruddlesden-Popper (RP) series of perovskite oxides, which is represented by the chemical formula of A$_{t+1}$B$_{t}$O$_{3t+1}$ (A and B being alkaline earth and transition metal elements, respectively), and their family compounds, where one or more conductive ABO$_3$ layers were sandwiched between AO insulating block layers \cite{PRB_91_Nozaki, PRL_08_Moon, PRB_11_Malvestuto}.  
These layered complex oxides, which are both low dimensional and strongly interacting, often exhibit unusual physical properties as a result of quantum fluctuation, such as high-$T_c$ superconductivity in cuprates \cite{RMP_98_Imada} and triplet superconductivity in ruthenates \cite{RMP_03_Mackenzie}.  
However, the synthesis of a homologous series of layered complex oxides is an extremely difficult task; in many cases, the number of conductive layer $t$ is up to 3 except for $t = \infty$, namely in the 3D limit \cite{PRB_91_Nozaki, PRL_08_Moon, PRB_11_Malvestuto}.  
Therefore, systematic control of the dimensionality from 3D to 2D with fixed fundamental electronic parameters has been exceptionally challenging and has never been reported.

Here, we propose a novel approach for investigating the quantum criticality using the dimensional crossover from 3D to 2D occurring in the quantum well (QW) structure of correlated oxides.  
In the QW structure, which has close structural similarities to the layered oxides but does not complicated interlayer interaction inherent in the layered oxides \cite{RMP_98_Imada, Nature_02_Valla, PRB_91_Nozaki, PRL_08_Moon, PRB_11_Malvestuto}, we digitally control the number of conductive layers of the strongly correlated oxide \cite{NatNanotech_14_King, Science_11_Yoshimatsu}.  
Furthermore, the strongly correlated electrons in the conductive layers are well confined within the potential well of the QW structures \cite{Science_11_Yoshimatsu, PRL_10_Yoshimatsu}.  
Consequently, the competition between the long-range order and the quantum fluctuation can be precisely investigated as a function of dimensionality (layer thickness $t$).

\begin{figure*}[t!]
\begin{center}
\includegraphics[width=15.5cm]{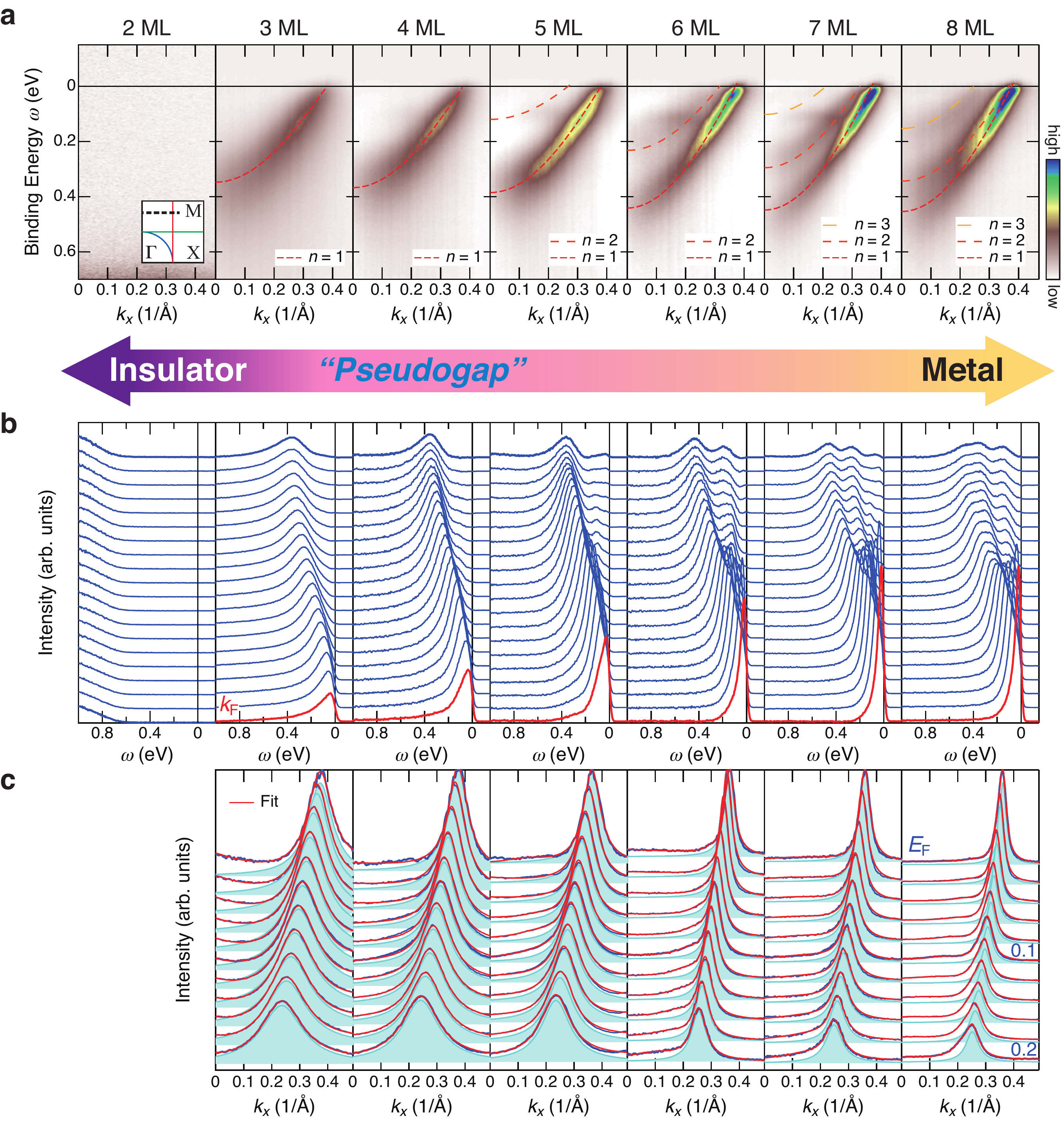}
\caption{A series of ARPES spectra of SrVO$_3$ quantum-well structures with various layer thicknesses. 
(a) Respective ARPES images of ultrathin SVO films with $t = 2$-8 ML. 
The ARPES data were acquired with the present photon energy of $h\nu = 88$ eV along a $k_x$ slice near the X point ($k_y = 0.75$$\pi/a$) as shown by the dashed line in the inset, which includes only the quantized $d_{zx}$ band.  
The intensity after subtracting the momentum-independent backgrounds at any point is given by a false color scale.  
Note that the series of ARPES spectra are normalized to the incident photon intensity, and the normalized intensity reflects the change in spectral weight of the QP states with changing $t$.  
The dashed lines show the results of the tight-binding fitting for each QW state.  
The inset shows the in-plane Fermi surface and the ARPES measured cut.  
(b) Corresponding EDCs to the respective ARPES images.  
The momentum-independent backgrounds have been subtracted from all the spectra.  
(c) Line-shape analysis for the MDCs at various $\omega$.  
The shaded areas indicate the Lorentzian functions for the $n = 1$ states.  
The $n = 1$ subband is prominent in comparison with the other subbands owing to the strong $h\nu$-dependent intensity modulation \cite{Science_11_Yoshimatsu, PRL_15_Kobayashi}.
}
\label{ARPES_SVO}
\end{center}
\end{figure*}

We have chosen ultrathin films of the conductive oxide SrVO$_3$ (SVO) epitaxially grown on Nb-doped SrTiO$_3$ substrates as the QW structures for this study \cite{Science_11_Yoshimatsu, PRL_10_Yoshimatsu, PRL_15_Kobayashi, PRL_12_Aizaki}.  
Bulk SVO is a typical FL metal with the simple configuration of $3d$ $t_{2g}$$^1$ \cite{PRL_12_Aizaki, PRB_98_Inoue}.  
The two-dimensional FL nature of the confined correlated electron in the QW structures has been observed down to $t = 6$ monolayers (ML) by \textit{in situ} angle-resolved photoemission spectroscopy (ARPES): The line shape of the ARPES spectra of quantized subbands is well described in the framework of the FL picture \cite{PRL_15_Kobayashi, PRL_12_Aizaki}.  
In addition, our previous angle-integrated photoemission (AIPES) study has revealed that the SVO-QW structures undergo a thickness-dependent transition from the FL metal to Mott insulator at a critical film thickness ($t_c$) of 2-3 ML via the “pseudogap” region of 3-5 ML \cite{PRL_10_Yoshimatsu, PRL_15_Kobayashi}.  
Thus, the SVO-QW structure is an ideal platform for investigating the quantum criticality induced by the enhancement of quantum fluctuation driven by dimensional crossover, as well as the change in two-dimensional FL states (QW states) on the borderline of a Mott insulating phase.

%\section{Experimental}
Digitally controlled SVO ultrathin films were grown on the atomically flat surface of TiO$_2$-terminated Nb-doped SrTiO$_3$(001) substrates in a laser molecular-beam epitaxy chamber connected to an ARPES system at BL-28 and BL-2A of the Photon Factory (PF) \cite{Science_11_Yoshimatsu, PRL_10_Yoshimatsu, PRL_15_Kobayashi, PRL_12_Aizaki}.  
During the growth of an SVO film, the thickness was precisely controlled on the atomic scale by monitoring the intensity oscillation of reflection high-energy electron diffraction (RHEED).  
The details of the growth conditions are described elsewhere \cite{Science_11_Yoshimatsu, PRL_10_Yoshimatsu}.

After growth, the samples were transferred to the photoemission chamber under an ultrahigh vacuum of 10$^{-10}$ Torr to avoid degradation of the SVO surface upon exposure to air.  
The ARPES experiments were conducted {\it in situ} at a temperature of 20 K using horizontal linear polarization of the incident light.  
The incident photon energy $h\nu$ was 88 eV, the photon momentum of which corresponds to the $\Gamma$ point along the surface normal direction in bulk SVO.  
The energy and angular resolutions were respectively set to about 30 meV and 0.3$^\circ$.  
The $E_\mathrm{F}$ of the samples was calibrated by measuring a gold foil that was electrically connected to the samples.  
The details of the ARPES measurement setups are described elsewhere \cite{Science_11_Yoshimatsu, PRL_15_Kobayashi}.

%\section{Results and discussion}
Figure~\ref{ARPES_SVO}(a) shows a series of ARPES images for the ultrathin SVO films with $t = 2$-8 ML.  
Because these band dispersions have been taken along the cut indicated by dashed line in the inset, the ARPES images consist of only the quantized $d_{zx}$ bands of V $3d$ $t_{2g}$ states with quantum numbers $n = 1$, 2, and 3 from the bottom \cite{Science_11_Yoshimatsu, PRL_15_Kobayashi}.  
Here, the series of ARPES images are normalized to the incident photon flux; hence, the color scale reflects the change in spectral weight as a function of $t$.  
The occurrence of the thickness-dependent metal-insulator transition (MIT) at $t_c$ of 2-3 ML is clearly observed as evidence of the disappearance of a QW subband(s) at 2 ML.  
Combined with the previous AIPES results \cite{PRL_10_Yoshimatsu}, the MIT is caused by the gradual disappearance of the subbands, while keeping the fundamental dispersion fixed, as a result of the spectral weight transfer from the coherent band (subband) near the Fermi level ($E_\mathrm{F}$) to the incoherent states located at 1.5 eV \cite{NatNanotech_14_King, PRL_10_Yoshimatsu}.

The anomalous spectral changes with approaching the Mott transition are further confirmed by the ARPES spectra near $E_\mathrm{F}$. 
We show the energy distribution curves (EDCs) and momentum distribution curves (MDCs) corresponding to the ARPES images in Figs.~\ref{ARPES_SVO}(b) and \ref{ARPES_SVO}(c), respectively.  
Hereafter, we focus our attention on the behavior of the $n = 1$ state to reveal how the spectral behavior changes toward the MIT, since the $n = 1$ subband is prominent in comparison with the other subbands \cite{Science_11_Yoshimatsu, PRL_15_Kobayashi}.  
As expected from the ARPES images in Fig.~\ref{ARPES_SVO}(a), sharp QP peaks exist in the vicinity of $E_\mathrm{F}$, and its intensity remains almost unchanged down to 6 ML.  
With further decreasing t, the QP peaks gradually reduce their spectral intensity in the range of $t = 3$-5 ML, and finally fades into an energy gap of 0.5-0.7 eV at 2 ML, the value of which is in good agreement with that in previous AIPES results \cite{PRL_10_Yoshimatsu}.  
The dramatic reduction of the QP weight at $E_\mathrm{F}$ with approaching the MIT suggests the strongly correlated nature of the QW states \cite{NatNanotech_14_King, PRL_10_Yoshimatsu, PRL_15_Zhong}.  
Furthermore, associated with the reduction in the QP intensity, the MDC width $\Delta k$ gradually increases, as can be seen in Fig.~\ref{ARPES_SVO}(c).  
These spectral behaviors imply the emergence of intriguing ground states in the neighborhood of the Mott transition.

\begin{figure}[!t]
\centering
\includegraphics[width=8.5cm]{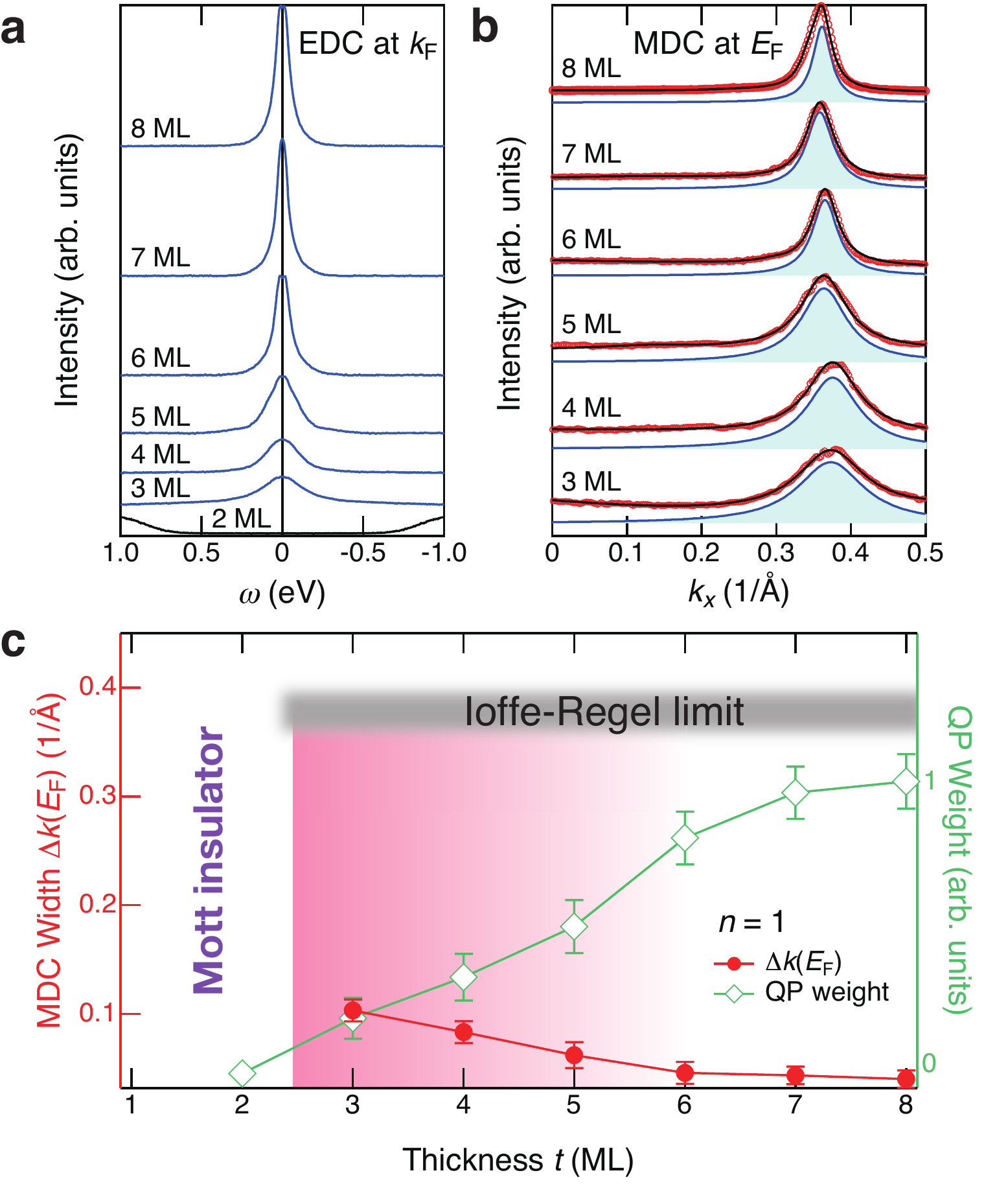}
\caption{Quasiparticle states at the Fermi level in SrVO$_3$ quantum-well structures. 
(a) QP peaks are obtained by the symmetrization of EDCs for $n = 1$ states at $k_\mathrm{F}$ [Fig.~\ref{ARPES_SVO}(b)] to $E_\mathrm{F}$.  
(b) MDCs at $E_\mathrm{F}$.  
Note that the intensities of the MDCs have been normalized to their peak height.  
The curves have been fitted to the combination of the Lorentzian function(s) corresponding to the respective subband(s) with a smooth background, and the Lorentzian functions of representative $n = 1$ states are displayed by blue-shaded areas.  
(c) Plot of MDC width $\Delta k (E_\mathrm{F})$ and QP weight for the $n = 1$ states with respect to $t$.  
The IR limit of corresponding states is also shown.  
The gradation area is the “pseudogap” (dimensional crossover) phase that is determined by previous AIPES \cite{PRL_10_Yoshimatsu}.  
Note that SVO-QW structures with $t \leq 2$ ML are Mott insulators.
}
\label{QP_SVOqw}
\end{figure}

In order to see the changes in more detail, we show symmetrized EDCs at the Fermi momentum $k_\mathrm{F}$ to $E_\mathrm{F}$ \cite{NatPhys_06_Kanigel} and MDCs at $E_\mathrm{F}$ in Figs.~\ref{QP_SVOqw}(a) and \ref{QP_SVOqw}(b), respectively.  
As shown in Fig.~\ref{QP_SVOqw}(a), it is clear that the QPs do not exhibit the pseudogap behaviors that are commonly observed in the ARPES spectra of underdoped high-$T_c$ cuprates \cite{NatPhys_06_Kanigel, NatMater_14_Hashimoto}.  
As $t$ decreases, the QP peak merely reduces its intensity, whereas its width is slightly broader.  
Simultaneously, $\Delta k$ at $E_\mathrm{F}$ [$\Delta k(E_\mathrm{F})$] also becomes broader, as summarized in the plot of Fig.~\ref{QP_SVOqw}(c).  
Although $\Delta k(E_\mathrm{F})$ slightly increases from 5 ML as t decreases toward the MIT, these values are well below the Ioffe-Regel (IR) limit of $k_\mathrm{F}/\Delta k(E_\mathrm{F}) \sim 1$ \cite{PS_60_Ioffe}, indicating that the effects of the disorder are not sufficiently strong to cause MIT in the present SVO-QW structures \cite{SciAdv_15_Mikheev, PRL_11_Scherwitzl}.  
Therefore, based on the ARPES-spectral analysis, it is naturally concluded that the thickness-dependent MIT is dominantly derived from the strong electron-electron correlation in the SVO-QW structures \cite{PRL_10_Yoshimatsu, PRL_15_Zhong}.

\begin{figure}[!t]
\centering
\includegraphics[width=8.0cm]{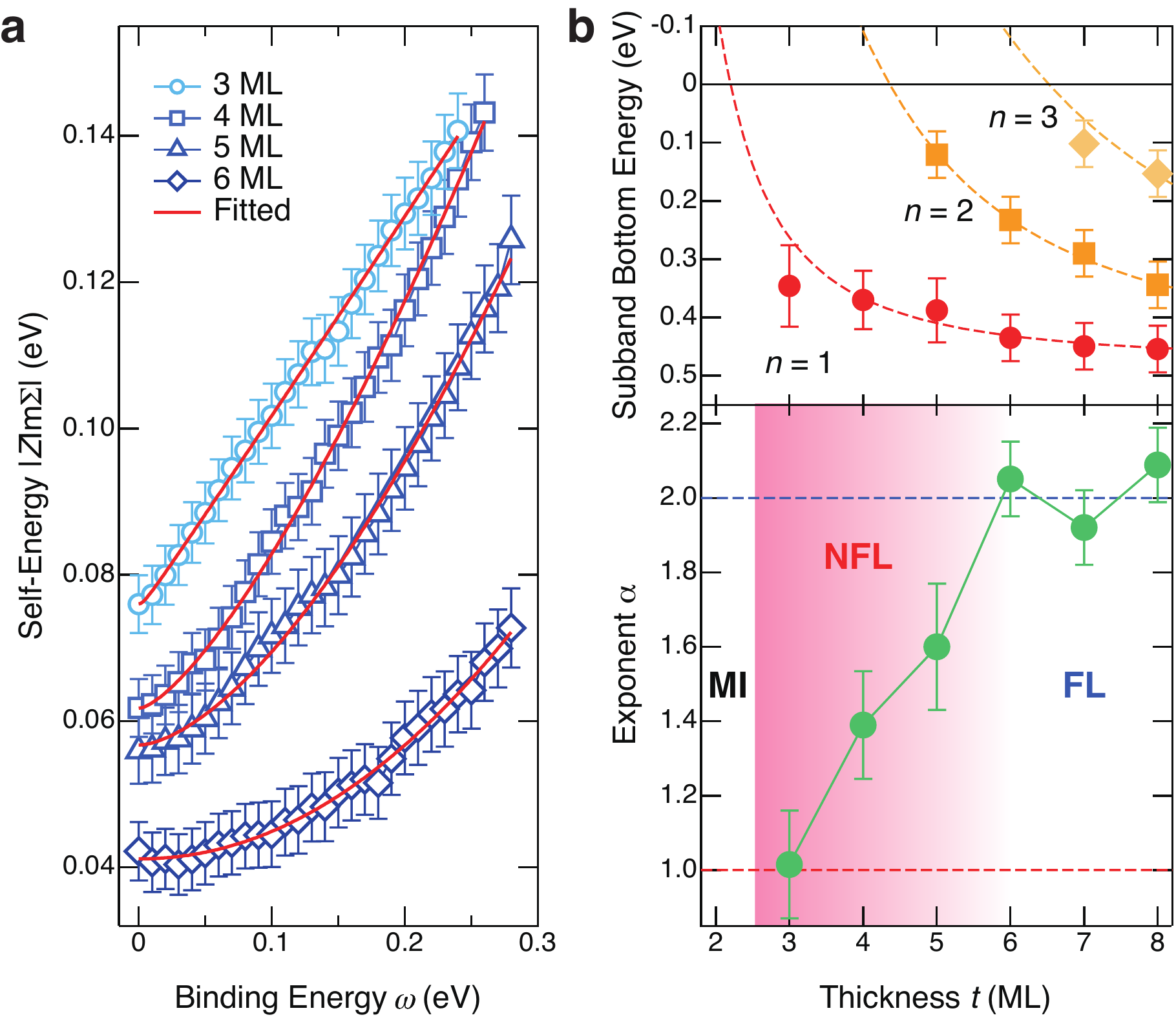}
\caption{Self-energy $\Sigma$ for SrVO$_3$ quantum-well states.  
(a) Imaginary part of the self-energy $\mathrm{Im}\Sigma$ for the $n = 1$ state as a function of $\omega$ for 3-6 ML.  
The solid curves represent the fitted curves based on Eq. (\ref{ZImS}).  
(b) (Top) Structure plot of the QW states as a function of $t$.  
With decreasing $t$, the number of subbands decreases, and eventually only a single QW subband appears in the two-dimensional limit of metallic SVO-QW structures (3 ML).  
The dashed lines represent the simulation results based on the phase shift quantization rule \cite{Science_11_Yoshimatsu}.  
The number of subbands in the QW states decreases with decreasing t in SVO-QW structures, and eventually only a single QW subband appears in the 3- or 4-ML SVO-QW structures.  
(Bottom) Plot of evaluated exponent $\alpha$ with respect to $t$.  
The value of $\alpha$ gradually reduces from 2 to 1 in the dimensional crossover region of 4-6 ML and then reaches to 1 at the two-dimensional limit of metallic QW states (3 ML) on the borderline of a Mott insulating phase.  
Here, FL, NFL, and MI denote Fermi liquid, non-Fermi liquid, and Mott-insulating states, respectively.  
The gradation area is the same as that in Fig.~\ref{QP_SVOqw}(c).
}
\label{ImS_SVOqw}
\end{figure}

The strange behavior of QP excitation is reminiscent of that in the vicinity of the quantum phase transition \cite{Nature_02_Valla}.  
In order to illuminate the underlying physics, we have evaluated the self-energy by employing the line-shape analysis of the MDCs as a function of the binding energy $\omega$ \cite{Science_99_Valla}.  
The obtained imaginary part of the self-energy $\mathrm{Im} \Sigma (\omega)$ for the $n = 1$ state of each QW structure is summarized in Fig.~\ref{ImS_SVOqw}(a).  
As can be seen in Fig.~\ref{ImS_SVOqw}(a), $\mathrm{Im} \Sigma (\omega)$ for 6 ML shows parabolic behavior, reflecting the FL ground states of the SVO-QW structure, as reported in previous studies \cite{PRL_15_Kobayashi, PRL_12_Aizaki}.  
These $\omega^2$ dependences of $\mathrm{Im} \Sigma (\omega)$ indicate that the correlated FL ground states of bulk SVO \cite{PRL_12_Aizaki, PRB_98_Inoue} hold in the region of $t \geq 6$ ML.  
However, when film thickness further approaches tc, the gradient of $\mathrm{Im} \Sigma (\omega)$ systematically changes from parabolic to linear.  
Eventually, the $\mathrm{Im} \Sigma (\omega)$ curve becomes linear at the thickness of 3 ML, which is the two-dimensional limit of the metallic SVO-QW structures.  
The linear $\omega$ dependence of $\mathrm{Im} \Sigma (\omega)$ at 3 ML is reminiscent of the marginal FL states in high-$T_c$ cuprates \cite{Science_99_Valla, PRL_89_Varma}.

These results indicate the occurrence of the crossover from FL to non-Fermi liquid (NFL) ground states in the vicinity of the MIT.  
To quantitatively address the crossover of the ground states, the $\mathrm{Im} \Sigma (\omega)$ curves are fitted to the following phenomenological form:
\begin{equation}
\left| Z \mathrm{Im} \Sigma \left( \omega \right) \right| = \Gamma^\mathrm{imp} + \beta' \left( \omega^2 + \left( \pi k_\mathrm{B}T \right)^2 \right) ^{\alpha/2},
\label{ZImS}
\end{equation}
where $Z$ is the renormalization factor, $\beta'$ denotes a coefficient reflecting the strength of the electron correlation, $k_\mathrm{B}$ is the Boltzmann constant, $\Gamma^\mathrm{imp}$ is the inverse lifetime of the QP associated with the impurity scattering, and $\alpha$ is the critical exponent.  
The fitting to Eq. (\ref{ZImS}) well reproduces the experimental $\mathrm{Im} \Sigma (\omega)$ curves, as shown by solid lines in Fig.~\ref{ImS_SVOqw}(a).  
The estimated values of $\alpha$ are plotted against t in Fig.~\ref{ImS_SVOqw}(b) together with the structure plot of SVO-QW states.  
The crossover from FL to NFL clearly occurs as $t$ is reduced from 6 ML.  
As $t$ approaches $t_c$, the value of $\alpha$ gradually reduces from 2 to 1 in the “pseudogap” (dimensional crossover) region of 4-6 ML \cite{PRL_10_Yoshimatsu} and then reaches 1 at 3 ML, indicating the existence of a quantum critical point (QCP) around $t_c$. 

From the structure plot of SVO-QW states in Fig.~\ref{ImS_SVOqw}(b), it is clear that the QCP emerges at the two-dimensional limit of the metallic QW structures (3 ML).  
Because only a single QW subband exits in the occupied states in the case of 3-4 ML, there is no intersubband interaction between the occupied states.  
Furthermore, the interference between the occupied $n = 1$ and unoccupied $n = 2$ states is expected to become negligibly weak at 3 ML owing to the largest energy separation between the two quantum states.  
Thus, in the two-dimensional limit of metallic SVO-QW structures, the QW states verge on the ideal two-dimensional states, suggesting that the QCP exists on the borderline between the metallic and Mott insulating phases as a result of the enhancement of quantum fluctuation in 2D.  
The close relationship between the emergence of the QCP and the energy diagram of the QW states suggests that the strength of the quantum fluctuation can be precisely controlled by tuning QW structures.

The present experiment strongly suggests the existence of QCP in the close proximity of the thickness-dependent Mott transition.  
Although the order parameter inducing the QCP is not clear at the moment, our observations have important implications in the search for novel quantum critical phenomena using metallic QW structures.  
The extraordinary physical properties of strongly correlated systems are usually found around a QCP, such as the ubiquitous formation of superconducting dome surrounding a QCP in the phase diagram of unconventional superconductors \cite{RMP_98_Imada, RMP_03_Mackenzie, PSSB_10_Sachdev, PNAS_10_Sebastian, Science_15_Ramshaw}.  
The artificially controllable QW structure of strongly correlated oxides with adjustable physical dimensions will provide a new strategy for designing the quantum critical phenomena emerging around a QCP.  
It has not escaped our notice that the quantum criticality observed in the two-dimensional limit of the metallic SVO-QW structure immediately suggests the possibility of superconductivity with optimal electron doping, since these SVO-QW structures are the mirror of hole-doped high-$T_c$ cuprates \cite{PRB_07_Arita}.

%\section{Acknowledgments}
The authors are very grateful to A. Santander-Syro, M. J. Rozenberg, T. Yoshida, and Y. Kuramoto for useful discussions. This work at KEK-PF was performed under the approval of the Program Advisory Committee (Proposals 2013S2-002, 2014G678, and 2015S2005) at the Institute of Materials Structure Science at KEK. This work was supported by a Grant-in-Aid for Scientific Research (B25287095, 15H02109, and 16H02115) and a Grant-in-Aid for Young Scientists (26870843) from the Japan Society for the Promotion of Science (JSPS), and the MEXT Elements Strategy Initiative to Form Core Research Center.

%references
%\bibliography{SVOqw_MIT}

%Figures

\end{document}